\documentclass[aps, pra, twocolumn, reprint, final, showpacs, superscriptaddress, floatfix, amssymb, amsmath, amsthm, hidelinks, numbers, sort&compress, floats]{revtex4-1}

\usepackage{epsfig}
\usepackage{graphicx}
\usepackage[normalem]{ulem}
\usepackage{xcolor}
\usepackage{times, graphics, dcolumn}
\usepackage{units}
\usepackage[colorlinks=true,citecolor=blue,allcolors=blue]{hyperref}  
\usepackage{bm, bbm}  
\usepackage[latin1]{inputenc}
\usepackage{ae,aecompl} 

\begin{document}

\title{Thermodynamic measurement of the sound velocity of a Bose gas across the transition to Bose-Einstein condensation}

\author{A R Fritsch}\email{amilson@usp.br}
\author{P E S Tavares}
\author{F A J Vivanco}
\author{G D Telles} 
\author{V S Bagnato}
\author{E A L Henn}
\affiliation{S\~{a}o Carlos Institute of Physics, University of S\~{a}o Paulo, PO Box 369, 13560-970, S\~{a}o Carlos, SP, Brazil}

\date{May 10, 2018}

\begin{abstract}
We present an alternative method for determining the sound velocity in atomic Bose-Einstein condensates, based on thermodynamic global variables. The total number of trapped atoms was as a function of temperature carefully studied across the phase transition, at constant volume. It allowed us to evaluate the sound velocity resulting in consistent values from the quantum to classical regime, in good agreement with previous results found in literature. We also provide some insight about the dominant sound mode (thermal or superfluid) across a wide temperature range. 
 \end{abstract}

\maketitle

\section{Introduction}

There are many striking differences between normal fluids, studied through standard hydrodynamics \cite{Landau1987Book}, and quantum fluids, which behavior is governed by quantum hydrodynamics \cite{Nozieres1999Book}. Some of them stand out such as the existence of two sound modes in superfluids \cite{Tisza1938} in contrast to a single, unique sound velocity in normal fluids. The origin of the two sound velocities in superfluids emerges from the motion of the superfluid component as an independent degree of freedom \cite{London1938,Landau1941}, eventually modeled, together with the normal component, in the two-fluid hydronamic theory \cite{Landau1941}.  This treatment has been developed and widely applied to liquid Helium and the seminal contributions are due to Tisza, Landau and Bogoliubov \cite{Landau1941,Tisza1940,Bogoliubov1947}.

The velocity of long wavelength excitations is directly connected to variations on the local pressure and density of the fluid. Indeed, as we will review later, the velocity of sound is proportional to the derivative of pressure with respect to density. In the two fluid model, since the superfluid component is treated separately from the normal component, there can be two independent modes for density perturbations: (i) the first sound, analogous to the sound velocity in normal fluids, where superfluid and normal components move in-phase, which is essentially a density wave and (ii) an out-of-phase oscillation of the components, also regarded as a temperature wave, named second-sound.

Although the theory has been developed for liquid Helium, it is not easy to experimentally investigate sound modes in this framework, since interactions are very strong. On the other hand, Bose-Einstein condensates, to which quantum hydrodynamics fully apply with spectacular results \cite{Jin1996}, are very versatile systems where the parameters (e.g. geometry, interactions, total number of particles, atomic density and temperature) can be tuned almost at will and, most importantly, the superfluid and normal components can be directly imaged and separately analyzed. Indeed, in atomic superfluids the first and second sound modes take a slightly different character: at high temperatures the former is identified as an oscillation of the thermal cloud density and the latter as an oscillation of the BEC density, while close to $T=0$ this is reverted.

The first measurements of the sound velocity in a BEC were performed by Andrews \textit{et al.} \cite{Andrews1997} by inducing a density perturbation in the BEC and observing its propagation as a function of time, allowing for a direct measurement of the excitation velocity. In order to compare their results with the theory, authors treated the density as homogeneous at half the BEC peak density, since they would not have access to local density or pressure of the fluid. This study triggered various  theoretical works \cite{Griffin1997A,Nikuni1998A} and eventually the sound velocity was studied in optical lattices \cite{Menotti2004,Ernst2010}, the excitation spectrum has been measured \cite{Steinhauer2002} and also shock waves were observed \cite{Damski2004,Simula2005}.

A few years ago, Meppelink \textit{et al.} \cite{Meppelink2009A} revisited and extended the early works by observing a sound wave in the BEC at finite temperatures and with improved techniques, allowing them to produce smaller density perturbations and therefore avoid non-linear effects. The obtained results display a good agreement with the theory. Recently, the sound modes of quantum fluids have been investigated in fermi gases \cite{Taylor2009A,Joseph2007,Sidorenkov2013}, lower dimensionality \cite{Luo2013}, in the context of spin-orbit coupled BECs \cite{Zheng:2012fk} and in the presence of disorder \cite{Gaul2009A}.

Here we take a different approach to investigate the sound velocity of a Bose gas at finite temperature across the Bose-Einstein condensation transition. In contrast to previous experimental investigations, we do not imprint a density perturbation in the BEC but we take a cloud in equilibrium and evaluate directly its global thermodinamic variables \cite{Romero-Rochin2012A}. By varying the number of atoms in the cloud for several different temperature and keeping the global volume parameter constant, we calculate the variation of the global pressure parameter with density and determine the sound velocity. This method allows us to compare our results to the original thermodynamic derivation of the sound velocity for a quantum gas. Besides, we obtain the contribution of both component separately for each temperature, which is not possible by the previous methods. 

\section{Theoretical Background}\label{sec:theory}

In the following we review the basic theoretical aspects related to the sound velocity in ultracold clouds and how we use the global thermodynamic variables to obtain it in this work. We restrict ourselves to key results and refer the reader to more complete and comprehensive reviews of these subjects \cite{PitaevskiiBook2003,PethickBook2008}. 

When studying small amplitude oscillations of the density of an uniform Bose gas in local thermal equilibrium, one can specify the local state of the system by its total mass density $\rho$, the superfluid velocity $v_s$, its temperature \textit{T} and the velocity of the excitations of the fluid.

Upon writing the continuity equation for the density and mass current one ends up with an equation relating the density and pressure $p$ of the fluid:
\begin{equation}
\frac{\partial^2\rho}{\partial t^2}=\nabla^2 p .
\label{pro}
\end{equation}

Also, one can write a relation between the velocity of the superfluid and the chemical potential $\mu$ 
which eventually leads, after some manipulation via thermodinamic relations, to
\begin{equation}
\frac{\partial^2 s}{\partial t^2}=\frac{\rho_s}{\rho}s^2\nabla^2 T ,
\label{sT}
\end{equation}
where $s$ is entropy per unit mass and $\rho_s$ is the superfluid density. Since $s$ depends on the density and temperature itself, one can immediately identify equations~(\ref{pro}) and (\ref{sT}) as coupled equations for temperature and pressure.

Solving the coupled equation by looking for small amplitude plane wave solutions one find a quartic equation for the velocity of the excitations, which, for $\rho_s\neq 0$ has two distinct solutions, identified as the first and second sound modes  of the quantum fluid. 

The sound modes have rather simple expressions at very specific limits. At very low temperatures, 
\begin{equation}
c_1^2=\frac{\partial p}{\partial \rho}
\label{c1}
\end{equation}
approaching the zero temperature Bogoliubov limit \cite{PethickBook2008}
\begin{equation}
c_1^2=\frac{gn}{m},
\label{t0}
\end{equation}
where $n$ is the number density and $g=\frac{4\pi\hbar^2 a}{m}$ with $a$ the s-wave scattering length. 

At higher temperatures with non-zero superfluid density, $c_1$ is composed by two terms: the first, similar to equation~(\ref{c1}), taken at constant temperature, and the second dependent on the entropy of the system \cite{PitaevskiiBook2003}. As we shall see later, we do not know how to explicity calculate the entropy of our system and, in this regime, we can only evaluate it at constant temperature. In any case, $c^2_1$ displays a linear behavior just before and across the critical temperature $T_c$. If $\rho_s=0$, there is a single solution for the sound velocity equation, which results in the usual sound velocity of a fluid  
\begin{equation}
c^2=\frac{\partial p}{\partial \rho}.
\end{equation}

Global thermodynamic variables were defined by Romero-Rochín \cite{Romero-RochinPRL2005} as an alternative approach to treat the thermodynamics of a gas which is not confined by rigid walls but rather by a non-homogeneous trap that extends everywhere in space and interacts with the atomic distribution differently at each point. We have been successfully using this approach to investigate different thermodynamic quantities in Bose gases \cite{Shiozaki2014A, PovedaCuevas2015A}.

In general, when dealing with non-homogeneous trapping potentials, the standard definitions of pressure (P) and volume (V) do not apply. Indeed, P and V are conjugate variables defined for homogeneous density distributions. In particular, P should have the same value in every position inside the volume occupied by the gas, in strong contrast to a non-uniform density distribution, where it changes from point-to-point in space. It is possible then to define the so-called global variables to describe the thermodynamics of an inhomogeneous system \cite{Romero-RochinBJP2005}. 

In brief, starting from thermodynamic and statistical mechanics assumptions we define a volume parameter and a pressure parameter, respectively:
\begin{equation}
\mathcal{V}=\frac{1}{\omega_x\omega_y\omega_z}
\end{equation}
and
\begin{eqnarray} 
\Pi &= \frac{2}{3\mathcal{V}}\langle U(\mathbf{r})\rangle ,\\
     &= \frac{2}{3\mathcal{V}}\int n(\mathbf{r})(\omega_x^2 x^2+\omega_y^2 y^2+\omega_z^2  z^2) d^3{r},\label{pressure}
\end{eqnarray}
where $\omega_i$ are the trapping frequencies, $\langle U(\mathbf{r})\rangle$ is the spatial mean of the external potential and $n(\mathbf{r})$ is the density distribution of the gas. It should be noted that, since we can write and experimentally identify $n(\mathbf{r})=n_{thermal}(\mathbf{r})+n_{BEC}(\mathbf{r})$, we can always write $\Pi=\Pi_{thermal}+\Pi_{BEC}$ and evaluate independently the pressure parameter of both components.

These variables have being proven to be a pair of conjugate variables, adequate to describe a non-homegeneous system, leading to the standard P and V in the thermodynamic limit \cite{Romero-Rochin2012A,Romero-RochinPRL2005,Silva2006}.

Under this context, we define a global first sound mode velocity for a Bose gas as:
\begin{equation}
c^2_{1g}=\frac{\partial \Pi}{\partial \rho} .
\label{1gs}
\end{equation}

Upon writing $\rho=\frac{mN}{\mathcal{V}}$, where $N$ is the total number of atoms each with mass $m$ confined in a Volume $\mathcal{V}$, with the volume kept fixed (as it is the case in our experiment, since we do not change the trap) equation \ref{1gs} leads to
\begin{equation}\label{eq:firstsoundlow3}
c_{1g}^{2} = \frac{\mathcal{V}}{m}\frac{\partial{\Pi}}{\partial{N}}.
\end{equation}

By computing the variation of $\Pi$ with $N$ in our experiment, we can immediately evaluate the first sound of the cloud.

\section{Experimental sequence}
The experimental setup and procedure is thoroughly described in previous works \cite{Henn2009,HennBJP}. The starting point of our experiments is a sample of  $^{87}$Rb atoms spin-polarized in the $|F=2,m_{F}=+2\rangle$ hyperfine state in a harmonic magnetic trap with trapping frequencies given by $\omega_z=2\pi \times 21.1(1)\unit{Hz}$ and $\omega_r=2\pi \times 188.2(3)\unit{Hz}$. Sound velocity (equation~(\ref{eq:firstsoundlow3})) is evaluated above, around and below the critical temperature $T_c$ for Bose-Einstein condensation. Atom-number ranges from $1 \times 10^6$ at $\unit[2]{\mu K}$ in a fully thermal cloud to $1-2 \times 10^{5}$ Bose-condensed atoms with no distinguishable thermal component, indicating condensed fractions $N_0/N > 70\%$. After reaching a final given state, the trap is switched off and the atoms are allowed to freely expand for  $\unit[23]{ms}$, after which they are imaged through standard resonant absorption imaging.
	We fit the cloud images with well-stablished bimodal (gaussian + Thomas-Fermi) profiles to determine: the number of condensed atoms, number of thermal atoms, typical sizes, densities and the temperature of the cloud. 
	
	 The number of atoms as well as their final temperature is controlled by increasing or decreasing the initial number of atoms loaded in the trap and/or by the radio frequency  used in the process of forced evaporative cooling. 
	 
	For this work we run the experiment varying the number of atoms and the temperature in the widest range allowed by our experimental setup to obtain a precise measurement of the sound velocity. After this set of data is obtained, we group together images in the same temperature range within $\pm\unit[5]{nK}$.  
		
		In order to calculate the global pressure parameter $\Pi$, equation~(\ref{pressure}), we use the well stablished Castin-Dum regression \cite{Castin1996} to determine the \textit{in situ} dimensions of the cloud, its in-trap density distribution and, together with the knowledge of the trapping potential, the pressure parameter for each cloud. This is the exact same procedure we have used in several previous works \cite{Romero-Rochin2012A,Shiozaki2014A,PovedaCuevas2015A} to evaluate the equation of state, the observation of a BEC, its compressibility, among other phenomena.
		
\begin{figure}[!ht]
	\centering
	\includegraphics[width=\linewidth]{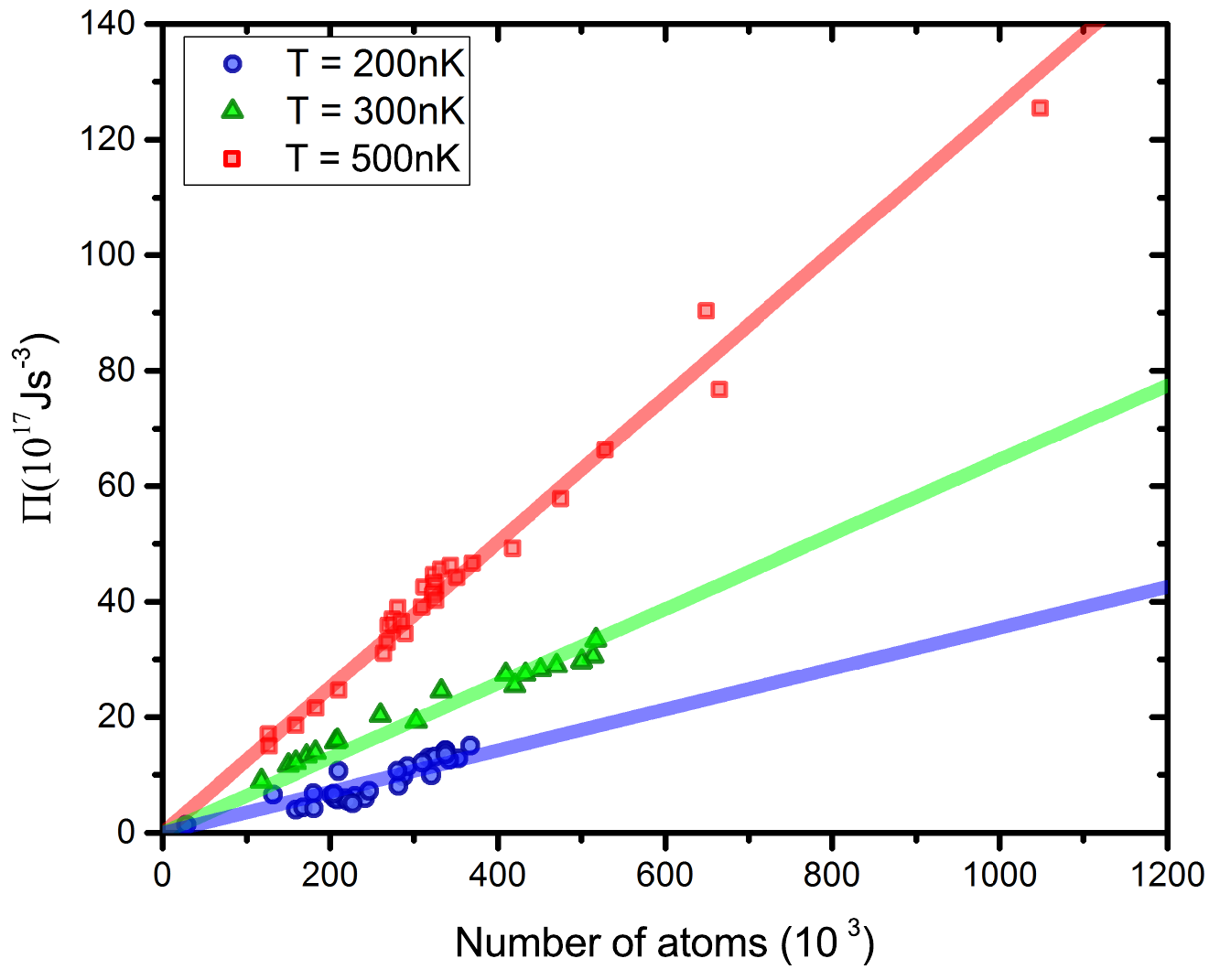}
	\caption{Typical sets of data used for calculating sound velocity. The squares are data for $\unit[500]{nK}$, triangles are data for $\unit[300]{nK}$ and circles are the selected data in the temperature of $\unit[200]{nK}$.}
	\label{fig:PivsN}
\end{figure}

		In figure~\ref{fig:PivsN} we show typical data for the pressure parameter obtained by the above-described procedure as a function of the number of atoms for three given temperatures. The displayed behavior is clearly linear and we extract the slope of this curve, which is the important quantity to evaluate the global sound velocity, by fitting it with a simple linear function. Tipically, we impose a minimum of 5 datapoints for each given temperature to confidently evaluate the data, but most of the datasets are much larger and comparable to what is shown in figure~\ref{fig:PivsN}.
		
		The sound velocity in our approach only depends on the slope of this plot, the mass of Rubidium atoms and the volume parameter, which for our experiment is $\mathcal{V} = \unit[5.4 \times 10^{-9}]{s^{3}}$ and remains the same since it only depends on the trapping frequencies.
		
\section{Results}

\begin{figure}
	\centering
	\includegraphics[width=\linewidth]{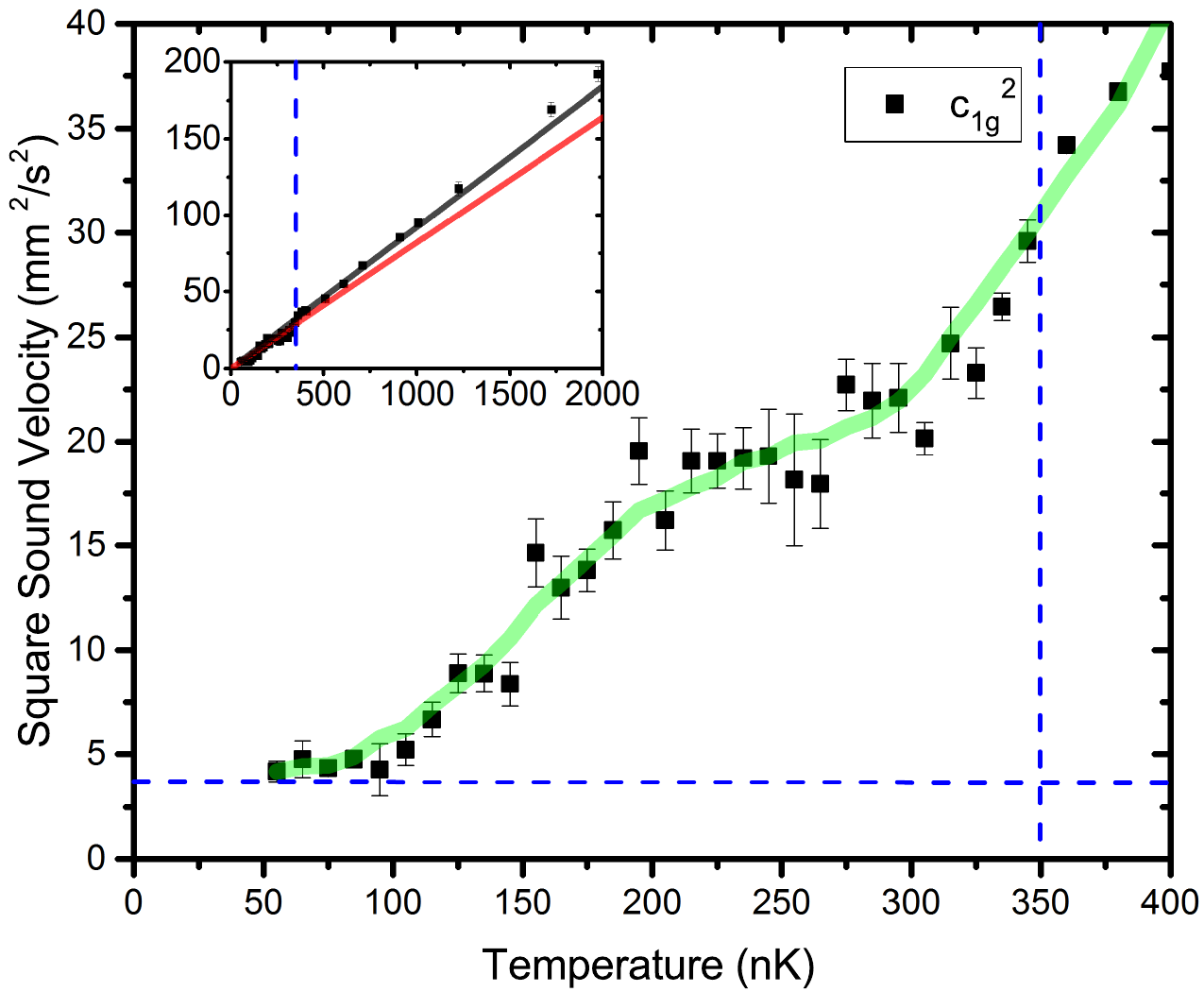}
	\caption{Squared (global) first sound mode of a Bose-Einstein condensate as a function of temperature. Solid lines are guides to the eye. Dashed horizontal line is the theoretical zero temperature value of $c_{1}^2$ (see text) and dashed vertical line is the approximate temperature above which no superfluid fraction is detected. Inset shows a broader range of temperatures for $c_{1g}^2$ where one can see the expected linear behavior for higher temperatures ($T>\unit[350]{nK}$). The red solid line in the inset presents the theoretical value for the sound velocity in thermal atoms and the black solid line is a linear fitting using the experimental data.}
	\label{fig:c1c2}
\end{figure}

	Figure~\ref{fig:c1c2} shows the squared first sound velocity, $c_{1g}^2$ as defined in equation~(\ref{eq:firstsoundlow3}), as a function of temperature evaluated by the procedure previously described. The first feature to be noticed is that at very low temperatures, the first sound mode approaches its zero temperature value, equation~(\ref{t0}), represented by the horizontal dashed line. Also, at higher temperatures and while $\rho_s\neq 0$, represented by the region of temperatures below the vertical dashed line, one observe the expected linear behavior of the first sound mode with temperature.
	
	The inset of figure~\ref{fig:c1c2} displays a wider range of temperatures, well into the region where $\rho_s = 0$. The observed behavior of the sound velocity in the thermal cloud is the expected linear growth which extrapolates to zero at $T=0$ in the absence of a BEC transition with a rate $\frac{\Delta \left( c_{1g}^2\right)}{\Delta T}=\unit[9.2(1)\times10^{-2}]{mm^2 s^{-2} nK^{-1}}$ (black line) close to the theoretical expected value $\unit[8.2\times10^{-2}]{mm^2 s^{-2} nK^{-1}}$ (red line). This small difference may be due to the other contribution for the sound that should be evaluated at constant entropy as already discussed in Section~\ref{sec:theory}. Besides quantitatively precisely, this also highlights an advantage of our method: one does not need to imprint density perturbations in the cloud to measure the sound velocity. Those would be very difficult to follow in a thermal cloud due to the natural damping, not allowing to measure any sound mode velocity in this regime. Our method, on its turn, allow to quantity the sound velocity with $\rho_s\neq 0$, $\rho_s= 0$ and across the BEC transition.
		
We can also obtain two other sets of quantities related to the sound velocity and which provide some additional insight to the results. The first quantities are the partial components of the first sound velocity, given by the independent derivative of the BEC and thermal pressure parameters with respect to the total number of atoms:
\begin{equation}
c_{BEC-part}^2 = \frac{\mathcal{V}}{m}\frac{\partial{\Pi_{BEC}}}{\partial{N}} \label{cBEC-partial}
\end{equation}
and
\begin{equation}
c_{ther-part}^2 = \frac{\mathcal{V}}{m}\frac{\partial{\Pi_{ther}}}{\partial{N}} . \label{cther-partial}
\end{equation}
One can immediately note that the defined global first sound velocity, equation~(\ref{1gs}), is given by the sum of the above-defined partial components
\begin{equation}
c_{1g}^2=c_{BEC-part}^2+c_{ther-part}^2 .
\end{equation}

The second set of quantities are the independent BEC and thermal sound velocities as if they were completely independent fluids:
\begin{equation}
c_{BEC}^2 = \frac{\mathcal{V}}{m}\frac{\partial{\Pi_{BEC}}}{\partial{N_{BEC}}} \label{c-BEC}
\end{equation}
and
\begin{equation}
c_{ther}^2 = \frac{\mathcal{V}}{m}\frac{\partial{\Pi_{ther}}}{\partial{N_{ther}}} . \label{c-ther} 
\end{equation}
Note that
\begin{equation}
c_{BEC-part}^2+c_{ther-part}^2 \neq c_{BEC}^2+c_{ther}^2,
\end{equation}
but still both can give valuable information as we discuss in the following.

\begin{figure}
	\centering
	\includegraphics[width=\linewidth]{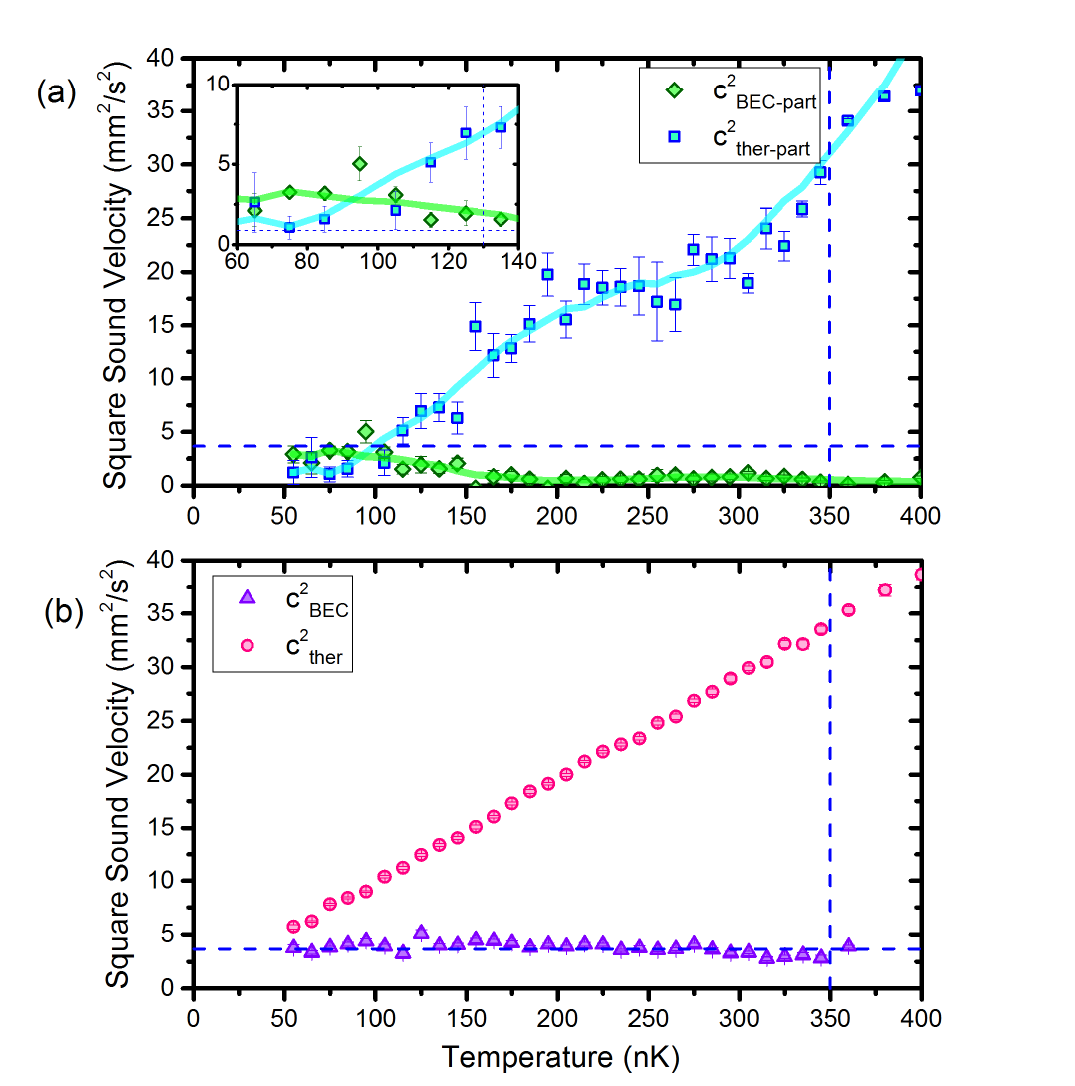}
	\caption{(a) Partial components of the first global sound velocity, given by the BEC (green diamonds) and thermal parts (blue squares).  (b) ``Single fluid'' (see text) sound velocity of the BEC (purple up-triangles) and thermal fluids (magenta circles). The solid lines in both (a) and (b) are guides to the eye following the points. The horizontal dashed line is the theoretical predicted $T=0$ first sound velocity and the vertical dashed line is the approximate limit above which no superfluid fraction is detected at our experiments.}
	\label{fig:cthcbec}
\end{figure}
Figure~\ref{fig:cthcbec} shows the partial components of the sound velocity. In both cases we plot solid lines just for eye guidance. In figure~\ref{fig:cthcbec}(a) one can see the range where each component dominates the first sound mode: below $\approx \unit[90]{nK}$ the BEC component dominates a vanishingly small thermal component while above it the thermal cloud dominates. The inset shows the lowest temperatures to emphasize the   
predominance of each component.

When we compute the independent sound velocities, as shown in figure~\ref{fig:cthcbec}(b), it is even more clear that the sound velocity in the thermal component varies linearly with the temperature all the way down to zero temperature. The propagation of the sound in the BEC part has a mean value of $c_{BEC}^2 = \unit[3.71(8)]{mm^2/s^2}$, found by a linear fitting with all these points, with a low dispersion around it. This value is represented by the horizontal blue dashed line. Evaluating the Bogoliubov sound velocity (equation~(\ref{t0})) that is valid only for pure BEC we obtain $c_{1}^2 = \unit[3.65(9)]{mm^2/s^2}$ if we use the four points at the lowest temperatures and $c_{1}^2 = \unit[3.82(5)]{mm^2/s^2}$ using all data for the BEC above the critical temperature. This last result shows that our method is very precise to determine the sound velocity in each component as if they were completely independent. Besides, since our measurements for the sound velocity in the BEC component shows very good agreement with the Bogoliubov theory, we conclude that the thermal part does not affect the sound propagation in the BEC at lower temperatures, which is expected, since the density of the thermal component is too low compared to the BEC density. As showed by the deviation from the predicted value, this velocity is affected only if we consider higher temperatures where the thermal component is dominant over the BEC.

In conclusion, we have defined and evaluated a global sound velocity of a Bose gas across a wide range of temperatures down to the Bose-Einstein condensation transition. We have also shown that the BEC and thermal components change dominance and character with respect to the sound mode depending on the temperature. We have also shown that several extra qualitative and quantitative features are reproduced by our analysis.

\begin{acknowledgments}
This work was supported by FAPESP under the program CePID (2013/07276-1). A.R.F. acknowledge CAPES and P.E.S.T. acknowledge FAPESP (2017/15753-5) scholarships and EALH acknowledges support from FAPESP 2015/20475-9 and VSB acknowledges research fellowships.
\end{acknowledgments}

\bibliographystyle{apsrev4-1}
\bibliography{SoundVelocity-References}

\begin{thebibliography}{34}%
\makeatletter
\providecommand \@ifxundefined [1]{%
 \@ifx{#1\undefined}
}%
\providecommand \@ifnum [1]{%
 \ifnum #1\expandafter \@firstoftwo
 \else \expandafter \@secondoftwo
 \fi
}%
\providecommand \@ifx [1]{%
 \ifx #1\expandafter \@firstoftwo
 \else \expandafter \@secondoftwo
 \fi
}%
\providecommand \natexlab [1]{#1}%
\providecommand \enquote  [1]{``#1''}%
\providecommand \bibnamefont  [1]{#1}%
\providecommand \bibfnamefont [1]{#1}%
\providecommand \citenamefont [1]{#1}%
\providecommand \href@noop [0]{\@secondoftwo}%
\providecommand \href [0]{\begingroup \@sanitize@url \@href}%
\providecommand \@href[1]{\@@startlink{#1}\@@href}%
\providecommand \@@href[1]{\endgroup#1\@@endlink}%
\providecommand \@sanitize@url [0]{\catcode `\\12\catcode `\$12\catcode
  `\&12\catcode `\#12\catcode `\^12\catcode `\_12\catcode `\%12\relax}%
\providecommand \@@startlink[1]{}%
\providecommand \@@endlink[0]{}%
\providecommand \url  [0]{\begingroup\@sanitize@url \@url }%
\providecommand \@url [1]{\endgroup\@href {#1}{\urlprefix }}%
\providecommand \urlprefix  [0]{URL }%
\providecommand \Eprint [0]{\href }%
\providecommand \doibase [0]{http://dx.doi.org/}%
\providecommand \selectlanguage [0]{\@gobble}%
\providecommand \bibinfo  [0]{\@secondoftwo}%
\providecommand \bibfield  [0]{\@secondoftwo}%
\providecommand \translation [1]{[#1]}%
\providecommand \BibitemOpen [0]{}%
\providecommand \bibitemStop [0]{}%
\providecommand \bibitemNoStop [0]{.\EOS\space}%
\providecommand \EOS [0]{\spacefactor3000\relax}%
\providecommand \BibitemShut  [1]{\csname bibitem#1\endcsname}%
\let\auto@bib@innerbib\@empty
\bibitem [{\citenamefont {Landau}\ and\ \citenamefont
  {Lifshitz}(1987)}]{Landau1987Book}%
  \BibitemOpen
  \bibfield  {author} {\bibinfo {author} {\bibfnamefont {L.~D.}\ \bibnamefont
  {Landau}}\ and\ \bibinfo {author} {\bibfnamefont {E.~M.}\ \bibnamefont
  {Lifshitz}},\ }\href {\doibase 10.1002/qj.49710444026} {\emph {\bibinfo
  {title} {Fluid Mechanics}}},\ Course of Theoretical Physics\ (\bibinfo
  {publisher} {Pergamon press},\ \bibinfo {year} {1987})\BibitemShut {NoStop}%
\bibitem [{\citenamefont {Nozi\`{e}res}\ and\ \citenamefont
  {Pines}(1999)}]{Nozieres1999Book}%
  \BibitemOpen
  \bibfield  {author} {\bibinfo {author} {\bibfnamefont {P.}~\bibnamefont
  {Nozi\`{e}res}}\ and\ \bibinfo {author} {\bibfnamefont {D.}~\bibnamefont
  {Pines}},\ }\href {\doibase 10.1016/0016-0032(66)90362-0} {\emph {\bibinfo
  {title} {The Theory of Quantum Liquids}}}\ (\bibinfo  {publisher} {Perseus
  Books Publishing},\ \bibinfo {year} {1999})\BibitemShut {NoStop}%
\bibitem [{\citenamefont {Tisza}(1938)}]{Tisza1938}%
  \BibitemOpen
  \bibfield  {author} {\bibinfo {author} {\bibfnamefont {L.}~\bibnamefont
  {Tisza}},\ }\href {\doibase 10.1038/141913a0} {\bibfield  {journal} {\bibinfo
   {journal} {Nature}\ }\textbf {\bibinfo {volume} {141}},\ \bibinfo {pages}
  {913} (\bibinfo {year} {1938})}\BibitemShut {NoStop}%
\bibitem [{\citenamefont {London}(1938)}]{London1938}%
  \BibitemOpen
  \bibfield  {author} {\bibinfo {author} {\bibfnamefont {F.}~\bibnamefont
  {London}},\ }\href {\doibase 10.1038/141643a0} {\bibfield  {journal}
  {\bibinfo  {journal} {Nature}\ }\textbf {\bibinfo {volume} {141}},\ \bibinfo
  {pages} {643} (\bibinfo {year} {1938})}\BibitemShut {NoStop}%
\bibitem [{\citenamefont {Landau}(1941)}]{Landau1941}%
  \BibitemOpen
  \bibfield  {author} {\bibinfo {author} {\bibfnamefont {L.}~\bibnamefont
  {Landau}},\ }\href {\doibase 10.1103/PhysRev.60.356} {\bibfield  {journal}
  {\bibinfo  {journal} {Phys. Rev.}\ }\textbf {\bibinfo {volume} {60}},\
  \bibinfo {pages} {356} (\bibinfo {year} {1941})}\BibitemShut {NoStop}%
\bibitem [{\citenamefont {Tisza}(1940)}]{Tisza1940}%
  \BibitemOpen
  \bibfield  {author} {\bibinfo {author} {\bibfnamefont {L.}~\bibnamefont
  {Tisza}},\ }\href@noop {} {\bibfield  {journal} {\bibinfo  {journal} {J.
  Phys. Radium}\ }\textbf {\bibinfo {volume} {1}},\ \bibinfo {pages} {164}
  (\bibinfo {year} {1940})}\BibitemShut {NoStop}%
\bibitem [{\citenamefont {Bogoliubov}(1947)}]{Bogoliubov1947}%
  \BibitemOpen
  \bibfield  {author} {\bibinfo {author} {\bibfnamefont {N.}~\bibnamefont
  {Bogoliubov}},\ }\href@noop {} {\bibfield  {journal} {\bibinfo  {journal} {J.
  Phys.}\ }\textbf {\bibinfo {volume} {11}},\ \bibinfo {pages} {23} (\bibinfo
  {year} {1947})}\BibitemShut {NoStop}%
\bibitem [{\citenamefont {Jin}\ \emph {et~al.}(1996)\citenamefont {Jin},
  \citenamefont {Ensher}, \citenamefont {Matthews}, \citenamefont {Wieman},\
  and\ \citenamefont {Cornell}}]{Jin1996}%
  \BibitemOpen
  \bibfield  {author} {\bibinfo {author} {\bibfnamefont {D.~S.}\ \bibnamefont
  {Jin}}, \bibinfo {author} {\bibfnamefont {J.~R.}\ \bibnamefont {Ensher}},
  \bibinfo {author} {\bibfnamefont {M.~R.}\ \bibnamefont {Matthews}}, \bibinfo
  {author} {\bibfnamefont {C.~E.}\ \bibnamefont {Wieman}}, \ and\ \bibinfo
  {author} {\bibfnamefont {E.~A.}\ \bibnamefont {Cornell}},\ }\href@noop {}
  {\bibfield  {journal} {\bibinfo  {journal} {Phys. Rev. Lett.}\ }\textbf
  {\bibinfo {volume} {77}},\ \bibinfo {pages} {420} (\bibinfo {year}
  {1996})}\BibitemShut {NoStop}%
\bibitem [{\citenamefont {Andrews}\ \emph {et~al.}(1997)\citenamefont
  {Andrews}, \citenamefont {Kurn}, \citenamefont {Miesner}, \citenamefont
  {Durfee}, \citenamefont {Townsend}, \citenamefont {Inouye},\ and\
  \citenamefont {Ketterle}}]{Andrews1997}%
  \BibitemOpen
  \bibfield  {author} {\bibinfo {author} {\bibfnamefont {M.~R.}\ \bibnamefont
  {Andrews}}, \bibinfo {author} {\bibfnamefont {D.~M.}\ \bibnamefont {Kurn}},
  \bibinfo {author} {\bibfnamefont {H.-J.}\ \bibnamefont {Miesner}}, \bibinfo
  {author} {\bibfnamefont {D.~S.}\ \bibnamefont {Durfee}}, \bibinfo {author}
  {\bibfnamefont {C.~G.}\ \bibnamefont {Townsend}}, \bibinfo {author}
  {\bibfnamefont {S.}~\bibnamefont {Inouye}}, \ and\ \bibinfo {author}
  {\bibfnamefont {W.}~\bibnamefont {Ketterle}},\ }\href {\doibase
  10.1103/PhysRevLett.79.553} {\bibfield  {journal} {\bibinfo  {journal} {Phys.
  Rev. Lett.}\ }\textbf {\bibinfo {volume} {79}},\ \bibinfo {pages} {553}
  (\bibinfo {year} {1997})}\BibitemShut {NoStop}%
\bibitem [{\citenamefont {Griffin}\ and\ \citenamefont
  {Zaremba}(1997)}]{Griffin1997A}%
  \BibitemOpen
  \bibfield  {author} {\bibinfo {author} {\bibfnamefont {A.}~\bibnamefont
  {Griffin}}\ and\ \bibinfo {author} {\bibfnamefont {E.}~\bibnamefont
  {Zaremba}},\ }\href {\doibase 10.1103/PhysRevA.56.4839} {\bibfield  {journal}
  {\bibinfo  {journal} {Phys. Rev. A}\ }\textbf {\bibinfo {volume} {56}},\
  \bibinfo {pages} {4839} (\bibinfo {year} {1997})}\BibitemShut {NoStop}%
\bibitem [{\citenamefont {Nikuni}\ and\ \citenamefont
  {Griffin}(1998)}]{Nikuni1998A}%
  \BibitemOpen
  \bibfield  {author} {\bibinfo {author} {\bibfnamefont {T.}~\bibnamefont
  {Nikuni}}\ and\ \bibinfo {author} {\bibfnamefont {A.}~\bibnamefont
  {Griffin}},\ }\href {\doibase 10.1103/PhysRevA.58.4044} {\bibfield  {journal}
  {\bibinfo  {journal} {Phys. Rev. A}\ }\textbf {\bibinfo {volume} {58}},\
  \bibinfo {pages} {4044} (\bibinfo {year} {1998})}\BibitemShut {NoStop}%
\bibitem [{\citenamefont {Menotti}\ \emph {et~al.}(2004)\citenamefont
  {Menotti}, \citenamefont {Kr\"amer}, \citenamefont {Smerzi}, \citenamefont
  {Pitaevskii},\ and\ \citenamefont {Stringari}}]{Menotti2004}%
  \BibitemOpen
  \bibfield  {author} {\bibinfo {author} {\bibfnamefont {C.}~\bibnamefont
  {Menotti}}, \bibinfo {author} {\bibfnamefont {M.}~\bibnamefont {Kr\"amer}},
  \bibinfo {author} {\bibfnamefont {A.}~\bibnamefont {Smerzi}}, \bibinfo
  {author} {\bibfnamefont {L.}~\bibnamefont {Pitaevskii}}, \ and\ \bibinfo
  {author} {\bibfnamefont {S.}~\bibnamefont {Stringari}},\ }\href {\doibase
  10.1103/PhysRevA.70.023609} {\bibfield  {journal} {\bibinfo  {journal} {Phys.
  Rev. A}\ }\textbf {\bibinfo {volume} {70}},\ \bibinfo {pages} {023609}
  (\bibinfo {year} {2004})}\BibitemShut {NoStop}%
\bibitem [{\citenamefont {Ernst}\ \emph {et~al.}(2010)\citenamefont {Ernst},
  \citenamefont {G\"{o}tze}, \citenamefont {Krauser}, \citenamefont {Pyka},
  \citenamefont {L\"{u}hmann}, \citenamefont {Pfannkuche},\ and\ \citenamefont
  {Sengstock}}]{Ernst2010}%
  \BibitemOpen
  \bibfield  {author} {\bibinfo {author} {\bibfnamefont {P.~T.}\ \bibnamefont
  {Ernst}}, \bibinfo {author} {\bibfnamefont {S.}~\bibnamefont {G\"{o}tze}},
  \bibinfo {author} {\bibfnamefont {J.~S.}\ \bibnamefont {Krauser}}, \bibinfo
  {author} {\bibfnamefont {K.}~\bibnamefont {Pyka}}, \bibinfo {author}
  {\bibfnamefont {D.-S.}\ \bibnamefont {L\"{u}hmann}}, \bibinfo {author}
  {\bibfnamefont {D.}~\bibnamefont {Pfannkuche}}, \ and\ \bibinfo {author}
  {\bibfnamefont {K.}~\bibnamefont {Sengstock}},\ }\href {\doibase
  10.1038/nphys1476} {\bibfield  {journal} {\bibinfo  {journal} {Nature
  Physics}\ }\textbf {\bibinfo {volume} {6}},\ \bibinfo {pages} {56} (\bibinfo
  {year} {2010})}\BibitemShut {NoStop}%
\bibitem [{\citenamefont {Steinhauer}\ \emph {et~al.}(2002)\citenamefont
  {Steinhauer}, \citenamefont {Ozeri}, \citenamefont {Katz},\ and\
  \citenamefont {Davidson}}]{Steinhauer2002}%
  \BibitemOpen
  \bibfield  {author} {\bibinfo {author} {\bibfnamefont {J.}~\bibnamefont
  {Steinhauer}}, \bibinfo {author} {\bibfnamefont {R.}~\bibnamefont {Ozeri}},
  \bibinfo {author} {\bibfnamefont {N.}~\bibnamefont {Katz}}, \ and\ \bibinfo
  {author} {\bibfnamefont {N.}~\bibnamefont {Davidson}},\ }\href {\doibase
  10.1103/PhysRevLett.88.120407} {\bibfield  {journal} {\bibinfo  {journal}
  {Phys. Rev. Lett.}\ }\textbf {\bibinfo {volume} {88}},\ \bibinfo {pages}
  {120407} (\bibinfo {year} {2002})}\BibitemShut {NoStop}%
\bibitem [{\citenamefont {Damski}(2004)}]{Damski2004}%
  \BibitemOpen
  \bibfield  {author} {\bibinfo {author} {\bibfnamefont {B.}~\bibnamefont
  {Damski}},\ }\href {\doibase 10.1103/PhysRevA.69.043610} {\bibfield
  {journal} {\bibinfo  {journal} {Phys. Rev. A}\ }\textbf {\bibinfo {volume}
  {69}},\ \bibinfo {pages} {043610} (\bibinfo {year} {2004})}\BibitemShut
  {NoStop}%
\bibitem [{\citenamefont {Simula}\ \emph {et~al.}(2005)\citenamefont {Simula},
  \citenamefont {Engels}, \citenamefont {Coddington}, \citenamefont
  {Schweikhard}, \citenamefont {Cornell},\ and\ \citenamefont
  {Ballagh}}]{Simula2005}%
  \BibitemOpen
  \bibfield  {author} {\bibinfo {author} {\bibfnamefont {T.~P.}\ \bibnamefont
  {Simula}}, \bibinfo {author} {\bibfnamefont {P.}~\bibnamefont {Engels}},
  \bibinfo {author} {\bibfnamefont {I.}~\bibnamefont {Coddington}}, \bibinfo
  {author} {\bibfnamefont {V.}~\bibnamefont {Schweikhard}}, \bibinfo {author}
  {\bibfnamefont {E.~A.}\ \bibnamefont {Cornell}}, \ and\ \bibinfo {author}
  {\bibfnamefont {R.~J.}\ \bibnamefont {Ballagh}},\ }\href {\doibase
  10.1103/PhysRevLett.94.080404} {\bibfield  {journal} {\bibinfo  {journal}
  {Phys. Rev. Lett.}\ }\textbf {\bibinfo {volume} {94}},\ \bibinfo {pages}
  {080404} (\bibinfo {year} {2005})}\BibitemShut {NoStop}%
\bibitem [{\citenamefont {Meppelink}\ \emph {et~al.}(2009)\citenamefont
  {Meppelink}, \citenamefont {Koller},\ and\ \citenamefont {van~der
  Straten}}]{Meppelink2009A}%
  \BibitemOpen
  \bibfield  {author} {\bibinfo {author} {\bibfnamefont {R.}~\bibnamefont
  {Meppelink}}, \bibinfo {author} {\bibfnamefont {S.~B.}\ \bibnamefont
  {Koller}}, \ and\ \bibinfo {author} {\bibfnamefont {P.}~\bibnamefont {van~der
  Straten}},\ }\href {\doibase 10.1103/PhysRevA.80.043605} {\bibfield
  {journal} {\bibinfo  {journal} {Phys. Rev. A}\ }\textbf {\bibinfo {volume}
  {80}},\ \bibinfo {pages} {043605} (\bibinfo {year} {2009})}\BibitemShut
  {NoStop}%
\bibitem [{\citenamefont {Taylor}\ \emph {et~al.}(2009)\citenamefont {Taylor},
  \citenamefont {Hu}, \citenamefont {Liu}, \citenamefont {Pitaevskii},
  \citenamefont {Griffin},\ and\ \citenamefont {Stringari}}]{Taylor2009A}%
  \BibitemOpen
  \bibfield  {author} {\bibinfo {author} {\bibfnamefont {E.}~\bibnamefont
  {Taylor}}, \bibinfo {author} {\bibfnamefont {H.}~\bibnamefont {Hu}}, \bibinfo
  {author} {\bibfnamefont {X.-J.}\ \bibnamefont {Liu}}, \bibinfo {author}
  {\bibfnamefont {L.~P.}\ \bibnamefont {Pitaevskii}}, \bibinfo {author}
  {\bibfnamefont {A.}~\bibnamefont {Griffin}}, \ and\ \bibinfo {author}
  {\bibfnamefont {S.}~\bibnamefont {Stringari}},\ }\href {\doibase
  10.1103/PhysRevA.80.053601} {\bibfield  {journal} {\bibinfo  {journal} {Phys.
  Rev. A}\ }\textbf {\bibinfo {volume} {80}},\ \bibinfo {pages} {053601}
  (\bibinfo {year} {2009})}\BibitemShut {NoStop}%
\bibitem [{\citenamefont {Joseph}\ \emph {et~al.}(2007)\citenamefont {Joseph},
  \citenamefont {Clancy}, \citenamefont {Luo}, \citenamefont {Kinast},
  \citenamefont {Turlapov},\ and\ \citenamefont {Thomas}}]{Joseph2007}%
  \BibitemOpen
  \bibfield  {author} {\bibinfo {author} {\bibfnamefont {J.}~\bibnamefont
  {Joseph}}, \bibinfo {author} {\bibfnamefont {B.}~\bibnamefont {Clancy}},
  \bibinfo {author} {\bibfnamefont {L.}~\bibnamefont {Luo}}, \bibinfo {author}
  {\bibfnamefont {J.}~\bibnamefont {Kinast}}, \bibinfo {author} {\bibfnamefont
  {A.}~\bibnamefont {Turlapov}}, \ and\ \bibinfo {author} {\bibfnamefont
  {J.~E.}\ \bibnamefont {Thomas}},\ }\href {\doibase
  10.1103/PhysRevLett.98.170401} {\bibfield  {journal} {\bibinfo  {journal}
  {Phys. Rev. Lett.}\ }\textbf {\bibinfo {volume} {98}},\ \bibinfo {pages}
  {170401} (\bibinfo {year} {2007})}\BibitemShut {NoStop}%
\bibitem [{\citenamefont {Sidorenkov}\ \emph {et~al.}(2013)\citenamefont
  {Sidorenkov}, \citenamefont {Tey}, \citenamefont {Grimm}, \citenamefont
  {Hou}, \citenamefont {Pitaevskii},\ and\ \citenamefont
  {Stringari}}]{Sidorenkov2013}%
  \BibitemOpen
  \bibfield  {author} {\bibinfo {author} {\bibfnamefont {L.~A.}\ \bibnamefont
  {Sidorenkov}}, \bibinfo {author} {\bibfnamefont {M.~K.}\ \bibnamefont {Tey}},
  \bibinfo {author} {\bibfnamefont {R.}~\bibnamefont {Grimm}}, \bibinfo
  {author} {\bibfnamefont {Y.-H.}\ \bibnamefont {Hou}}, \bibinfo {author}
  {\bibfnamefont {L.}~\bibnamefont {Pitaevskii}}, \ and\ \bibinfo {author}
  {\bibfnamefont {S.}~\bibnamefont {Stringari}},\ }\href {\doibase
  10.1038/nature12136} {\bibfield  {journal} {\bibinfo  {journal} {Nature}\
  }\textbf {\bibinfo {volume} {498}},\ \bibinfo {pages} {78} (\bibinfo {year}
  {2013})}\BibitemShut {NoStop}%
\bibitem [{\citenamefont {Yun-Wen}\ and\ \citenamefont
  {Ji-Sheng}(2013)}]{Luo2013}%
  \BibitemOpen
  \bibfield  {author} {\bibinfo {author} {\bibfnamefont {L.}~\bibnamefont
  {Yun-Wen}}\ and\ \bibinfo {author} {\bibfnamefont {C.}~\bibnamefont
  {Ji-Sheng}},\ }\href {\doibase 10.1088/0253-6102/60/6/07} {\bibfield
  {journal} {\bibinfo  {journal} {Communications in Theoretical Physics}\
  }\textbf {\bibinfo {volume} {60}},\ \bibinfo {pages} {673} (\bibinfo {year}
  {2013})}\BibitemShut {NoStop}%
\bibitem [{\citenamefont {Zheng}\ and\ \citenamefont
  {Li}(2012)}]{Zheng:2012fk}%
  \BibitemOpen
  \bibfield  {author} {\bibinfo {author} {\bibfnamefont {W.}~\bibnamefont
  {Zheng}}\ and\ \bibinfo {author} {\bibfnamefont {Z.~B.}\ \bibnamefont {Li}},\
  }\href@noop {} {\bibfield  {journal} {\bibinfo  {journal} {Phys. Rev. A}\
  }\textbf {\bibinfo {volume} {85}},\ \bibinfo {pages} {053607} (\bibinfo
  {year} {2012})}\BibitemShut {NoStop}%
\bibitem [{\citenamefont {Gaul}\ \emph {et~al.}(2009)\citenamefont {Gaul},
  \citenamefont {Renner},\ and\ \citenamefont {M\"uller}}]{Gaul2009A}%
  \BibitemOpen
  \bibfield  {author} {\bibinfo {author} {\bibfnamefont {C.}~\bibnamefont
  {Gaul}}, \bibinfo {author} {\bibfnamefont {N.}~\bibnamefont {Renner}}, \ and\
  \bibinfo {author} {\bibfnamefont {C.~A.}\ \bibnamefont {M\"uller}},\ }\href
  {\doibase 10.1103/PhysRevA.80.053620} {\bibfield  {journal} {\bibinfo
  {journal} {Phys. Rev. A}\ }\textbf {\bibinfo {volume} {80}},\ \bibinfo
  {pages} {053620} (\bibinfo {year} {2009})}\BibitemShut {NoStop}%
\bibitem [{\citenamefont {Romero-Rochin}\ \emph {et~al.}(2012)\citenamefont
  {Romero-Rochin}, \citenamefont {Shiozaki}, \citenamefont {Caracanhas},
  \citenamefont {Henn}, \citenamefont {Magalh\~aes}, \citenamefont {Roati},\
  and\ \citenamefont {Bagnato}}]{Romero-Rochin2012A}%
  \BibitemOpen
  \bibfield  {author} {\bibinfo {author} {\bibfnamefont {V.}~\bibnamefont
  {Romero-Rochin}}, \bibinfo {author} {\bibfnamefont {R.~F.}\ \bibnamefont
  {Shiozaki}}, \bibinfo {author} {\bibfnamefont {M.}~\bibnamefont
  {Caracanhas}}, \bibinfo {author} {\bibfnamefont {E.~A.~L.}\ \bibnamefont
  {Henn}}, \bibinfo {author} {\bibfnamefont {K.~M.~F.}\ \bibnamefont
  {Magalh\~aes}}, \bibinfo {author} {\bibfnamefont {G.}~\bibnamefont {Roati}},
  \ and\ \bibinfo {author} {\bibfnamefont {V.~S.}\ \bibnamefont {Bagnato}},\
  }\href {\doibase 10.1103/PhysRevA.85.023632} {\bibfield  {journal} {\bibinfo
  {journal} {Phys. Rev. A}\ }\textbf {\bibinfo {volume} {85}},\ \bibinfo
  {pages} {023632} (\bibinfo {year} {2012})}\BibitemShut {NoStop}%
\bibitem [{\citenamefont {Pitaevskii}\ and\ \citenamefont
  {Stringari}(2003)}]{PitaevskiiBook2003}%
  \BibitemOpen
  \bibfield  {author} {\bibinfo {author} {\bibfnamefont {L.~P.}\ \bibnamefont
  {Pitaevskii}}\ and\ \bibinfo {author} {\bibfnamefont {S.}~\bibnamefont
  {Stringari}},\ }\href@noop {} {\emph {\bibinfo {title} {Bose-Einstein
  condensation}}}\ (\bibinfo  {publisher} {Oxford University Press},\ \bibinfo
  {address} {Oxford},\ \bibinfo {year} {2003})\ p.\ \bibinfo {pages}
  {382}\BibitemShut {NoStop}%
\bibitem [{\citenamefont {Pethick}\ and\ \citenamefont
  {Smith}(2008)}]{PethickBook2008}%
  \BibitemOpen
  \bibfield  {author} {\bibinfo {author} {\bibfnamefont {C.~J.}\ \bibnamefont
  {Pethick}}\ and\ \bibinfo {author} {\bibfnamefont {H.}~\bibnamefont
  {Smith}},\ }\href@noop {} {\emph {\bibinfo {title} {Bose-Einstein
  condensation in dilute gases}}},\ \bibinfo {number} {1991}\ (\bibinfo
  {publisher} {Cambridge University Press},\ \bibinfo {address} {2nd ed.
  Cambridge},\ \bibinfo {year} {2008})\ p.\ \bibinfo {pages} {569}\BibitemShut
  {NoStop}%
\bibitem [{\citenamefont {Romero-Roch\'{\i}n}(2005)}]{Romero-RochinPRL2005}%
  \BibitemOpen
  \bibfield  {author} {\bibinfo {author} {\bibfnamefont {V.}~\bibnamefont
  {Romero-Roch\'{\i}n}},\ }\href {\doibase 10.1103/PhysRevLett.94.130601}
  {\bibfield  {journal} {\bibinfo  {journal} {Phys. Rev. Lett.}\ }\textbf
  {\bibinfo {volume} {94}},\ \bibinfo {pages} {130601} (\bibinfo {year}
  {2005})}\BibitemShut {NoStop}%
\bibitem [{\citenamefont {Shiozaki}\ \emph {et~al.}(2014)\citenamefont
  {Shiozaki}, \citenamefont {Telles}, \citenamefont {Castilho}, \citenamefont
  {Poveda-Cuevas}, \citenamefont {Muniz}, \citenamefont {Roati}, \citenamefont
  {Romero-Rochin},\ and\ \citenamefont {Bagnato}}]{Shiozaki2014A}%
  \BibitemOpen
  \bibfield  {author} {\bibinfo {author} {\bibfnamefont {R.~F.}\ \bibnamefont
  {Shiozaki}}, \bibinfo {author} {\bibfnamefont {G.~D.}\ \bibnamefont
  {Telles}}, \bibinfo {author} {\bibfnamefont {P.}~\bibnamefont {Castilho}},
  \bibinfo {author} {\bibfnamefont {F.~J.}\ \bibnamefont {Poveda-Cuevas}},
  \bibinfo {author} {\bibfnamefont {S.~R.}\ \bibnamefont {Muniz}}, \bibinfo
  {author} {\bibfnamefont {G.}~\bibnamefont {Roati}}, \bibinfo {author}
  {\bibfnamefont {V.}~\bibnamefont {Romero-Rochin}}, \ and\ \bibinfo {author}
  {\bibfnamefont {V.~S.}\ \bibnamefont {Bagnato}},\ }\href {\doibase
  10.1103/PhysRevA.90.043640} {\bibfield  {journal} {\bibinfo  {journal} {Phys.
  Rev. A}\ }\textbf {\bibinfo {volume} {90}},\ \bibinfo {pages} {043640}
  (\bibinfo {year} {2014})}\BibitemShut {NoStop}%
\bibitem [{\citenamefont {Poveda-Cuevas}\ \emph {et~al.}(2015)\citenamefont
  {Poveda-Cuevas}, \citenamefont {Castilho}, \citenamefont {Mercado-Gutierrez},
  \citenamefont {Fritsch}, \citenamefont {Muniz}, \citenamefont {Lucioni},
  \citenamefont {Roati},\ and\ \citenamefont {Bagnato}}]{PovedaCuevas2015A}%
  \BibitemOpen
  \bibfield  {author} {\bibinfo {author} {\bibfnamefont {F.~J.}\ \bibnamefont
  {Poveda-Cuevas}}, \bibinfo {author} {\bibfnamefont {P.~C.~M.}\ \bibnamefont
  {Castilho}}, \bibinfo {author} {\bibfnamefont {E.~D.}\ \bibnamefont
  {Mercado-Gutierrez}}, \bibinfo {author} {\bibfnamefont {A.~R.}\ \bibnamefont
  {Fritsch}}, \bibinfo {author} {\bibfnamefont {S.~R.}\ \bibnamefont {Muniz}},
  \bibinfo {author} {\bibfnamefont {E.}~\bibnamefont {Lucioni}}, \bibinfo
  {author} {\bibfnamefont {G.}~\bibnamefont {Roati}}, \ and\ \bibinfo {author}
  {\bibfnamefont {V.~S.}\ \bibnamefont {Bagnato}},\ }\href {\doibase
  10.1103/PhysRevA.92.013638} {\bibfield  {journal} {\bibinfo  {journal} {Phys.
  Rev. A}\ }\textbf {\bibinfo {volume} {92}},\ \bibinfo {pages} {013638}
  (\bibinfo {year} {2015})}\BibitemShut {NoStop}%
\bibitem [{\citenamefont {Romero-Roch\'{\i}n}\ and\ \citenamefont
  {Bagnato}(2005)}]{Romero-RochinBJP2005}%
  \BibitemOpen
  \bibfield  {author} {\bibinfo {author} {\bibfnamefont {V.}~\bibnamefont
  {Romero-Roch\'{\i}n}}\ and\ \bibinfo {author} {\bibfnamefont {V.~S.}\
  \bibnamefont {Bagnato}},\ }\href {\doibase 10.1590/S0103-97332005000400004}
  {\bibfield  {journal} {\bibinfo  {journal} {{Brazilian Journal of Physics}}\
  }\textbf {\bibinfo {volume} {35}},\ \bibinfo {pages} {607} (\bibinfo {year}
  {2005})}\BibitemShut {NoStop}%
\bibitem [{\citenamefont {Silva}\ \emph {et~al.}(2006)\citenamefont {Silva},
  \citenamefont {Henn}, \citenamefont {Magalh{\~a}es}, \citenamefont
  {Marcassa}, \citenamefont {Romero-Rochin},\ and\ \citenamefont
  {Bagnato}}]{Silva2006}%
  \BibitemOpen
  \bibfield  {author} {\bibinfo {author} {\bibfnamefont {R.~R.}\ \bibnamefont
  {Silva}}, \bibinfo {author} {\bibfnamefont {E.~A.~L.}\ \bibnamefont {Henn}},
  \bibinfo {author} {\bibfnamefont {K.~M.~F.}\ \bibnamefont {Magalh{\~a}es}},
  \bibinfo {author} {\bibfnamefont {L.~G.}\ \bibnamefont {Marcassa}}, \bibinfo
  {author} {\bibfnamefont {V.}~\bibnamefont {Romero-Rochin}}, \ and\ \bibinfo
  {author} {\bibfnamefont {V.~S.}\ \bibnamefont {Bagnato}},\ }\href {\doibase
  10.1134/S1054660X06040244} {\bibfield  {journal} {\bibinfo  {journal} {Laser
  Physics}\ }\textbf {\bibinfo {volume} {16}},\ \bibinfo {pages} {687}
  (\bibinfo {year} {2006})}\BibitemShut {NoStop}%
\bibitem [{\citenamefont {Henn}\ \emph {et~al.}(2009)\citenamefont {Henn},
  \citenamefont {Seman}, \citenamefont {Roati}, \citenamefont {Magalh\~aes},\
  and\ \citenamefont {Bagnato}}]{Henn2009}%
  \BibitemOpen
  \bibfield  {author} {\bibinfo {author} {\bibfnamefont {E.~A.~L.}\
  \bibnamefont {Henn}}, \bibinfo {author} {\bibfnamefont {J.~A.}\ \bibnamefont
  {Seman}}, \bibinfo {author} {\bibfnamefont {G.}~\bibnamefont {Roati}},
  \bibinfo {author} {\bibfnamefont {K.~M.~F.}\ \bibnamefont {Magalh\~aes}}, \
  and\ \bibinfo {author} {\bibfnamefont {V.~S.}\ \bibnamefont {Bagnato}},\
  }\href {\doibase 10.1103/PhysRevLett.103.045301} {\bibfield  {journal}
  {\bibinfo  {journal} {Phys. Rev. Lett.}\ }\textbf {\bibinfo {volume} {103}},\
  \bibinfo {pages} {045301} (\bibinfo {year} {2009})}\BibitemShut {NoStop}%
\bibitem [{\citenamefont {Henn}\ \emph {et~al.}(2008)\citenamefont {Henn},
  \citenamefont {Seman}, \citenamefont {Seco}, \citenamefont {Olimpio},
  \citenamefont {Castilho}, \citenamefont {Roati}, \citenamefont {Magalh\~aes},
  \citenamefont {Magalh\~aes},\ and\ \citenamefont {Bagnato}}]{HennBJP}%
  \BibitemOpen
  \bibfield  {author} {\bibinfo {author} {\bibfnamefont {E.~A.~L.}\
  \bibnamefont {Henn}}, \bibinfo {author} {\bibfnamefont {J.~A.}\ \bibnamefont
  {Seman}}, \bibinfo {author} {\bibfnamefont {G.~B.}\ \bibnamefont {Seco}},
  \bibinfo {author} {\bibfnamefont {E.~P.}\ \bibnamefont {Olimpio}}, \bibinfo
  {author} {\bibfnamefont {P.}~\bibnamefont {Castilho}}, \bibinfo {author}
  {\bibfnamefont {G.}~\bibnamefont {Roati}}, \bibinfo {author} {\bibfnamefont
  {D.~V.}\ \bibnamefont {Magalh\~aes}}, \bibinfo {author} {\bibfnamefont
  {K.~M.~F.}\ \bibnamefont {Magalh\~aes}}, \ and\ \bibinfo {author}
  {\bibfnamefont {V.~S.}\ \bibnamefont {Bagnato}},\ }\href {\doibase
  10.1590/S0103-97332008000200012} {\bibfield  {journal} {\bibinfo  {journal}
  {{Brazilian Journal of Physics}}\ }\textbf {\bibinfo {volume} {38}},\
  \bibinfo {pages} {279 } (\bibinfo {year} {2008})}\BibitemShut {NoStop}%
\bibitem [{\citenamefont {Castin}\ and\ \citenamefont
  {Dum}(1996)}]{Castin1996}%
  \BibitemOpen
  \bibfield  {author} {\bibinfo {author} {\bibfnamefont {Y.}~\bibnamefont
  {Castin}}\ and\ \bibinfo {author} {\bibfnamefont {R.}~\bibnamefont {Dum}},\
  }\href {\doibase 10.1103/PhysRevLett.77.5315} {\bibfield  {journal} {\bibinfo
   {journal} {Phys. Rev. Lett.}\ }\textbf {\bibinfo {volume} {77}},\ \bibinfo
  {pages} {5315} (\bibinfo {year} {1996})}\BibitemShut {NoStop}%
\end{thebibliography}%

\end{document}